

\magnification = 1200

\font\titlerm = cmr10 scaled\magstep 4
\font\titlerms = cmr7 scaled\magstep 4
\font\titlermss = cmr5 scaled\magstep 4
\font\titlei = cmmi10 scaled\magstep 4
\font\titleis = cmmi7 scaled\magstep 4
\font\titleiss = cmmi5 scaled\magstep 4
\font\titlesy = cmsy10 scaled\magstep 4
\font\titlesys = cmsy7 scaled\magstep 4
\font\titlesyss = cmsy5 scaled\magstep 4
\font\titleit = cmti10 scaled\magstep 4

\def\titlefont{\def\rm{\fam0\titlerm}
\def\it{\fam\itfam\titleit}
\textfont0 = \titlerm
\scriptfont0 = \titlerms
\scriptscriptfont0 = \titlermss
\textfont1 = \titlei
\scriptfont1 = \titleis
\scriptscriptfont1 = \titleiss
\textfont2 = \titlesy
\scriptfont2 = \titlesys
\scriptscriptfont2 = \titlesyss
\textfont\itfam = \titleit
\rm}

\def\sectionfont{\def\rm{\fam0\tenrm}
\def\it{\fam\itfam\tenit}
\def\bf{\fam\bffam\tenbf}
\textfont0 = \tenrm
\scriptfont0 = \sevenrm
\scriptscriptfont0 = \fiverm
\textfont1 = \teni
\scriptfont1 = \seveni  \scriptscriptfont1=\fivei
\textfont2 = \tensy
\scriptfont2 = \sevensy
\scriptscriptfont2 = \fivesy
\textfont\itfam = \tenit
\textfont\bffam = \tenbf
\rm}

\font\teenyfont = cmr5

\global\baselineskip = 1.2\baselineskip
\global\parskip = 4pt plus 0.3pt
\global\nulldelimiterspace = 0pt

\predisplaypenalty 1000


\def\endignore{}
\def\ignore #1\endignore{}

\newcount\dflag
\dflag = 0


\def\monthname{\ifcase\month
\or Jan \or Feb \or Mar \or Apr \or May \or June%
\or July \or Aug \or Sept \or Oct \or Nov \or Dec
\fi}




\def\endid{}
\def\id#1\endid{\number\day\ \monthname \number\year
\hfill #1}

\def\endtitle{}
\def\title#1\endtitle{\vskip.15in\titlefont
\global\baselineskip = 2\baselineskip
#1\vskip.3in
\baselineskip = 0.5\baselineskip\sectionfont}

\def\lblfoot{This work was supported by the Director, Office of Energy
Research, Office of High Energy and Nuclear Physics, Division of High
Energy Physics of the U.S. Department of Energy under Contract
DE-AC03-76SF00098.}

\def\endauthors{}
\def\authors#1\endauthors{
#1\if\dflag = 0
\footnote{}{\noindent\lblfoot}\fi}

\def\endabstract{}
\def\abstract#1\endabstract{\vskip .2in%
\centerline{\sectionfont\bf Abstract}%
\vskip .1in%
\noindent#1%
\ifnum\dflag = 0
\footline = {\hfil}\pageno = 0
\vfill\eject
\pageno = 1\footline{\centerline{\sectionfont\folio}}
\fi\ifnum\dflag = 2
\footline = {\hfil}\pageno = 0
\vfill\eject
\fi}


\newcount\nsection
\newcount\nsubsection

\def\section#1{\global\advance\nsection by 1
\global\nsubsection = 0
\bigskip\noindent
\centerline{\sectionfont\bf\number\nsection.\ #1}
\nobreak\medskip\sectionfont\nobreak}

\def\subsection#1{\global\advance\nsubsection by 1
\bigskip\noindent
\centerline{\sectionfont \it \number\nsection.\number\nsubsection.\ #1}
\nobreak\smallskip\rm\nobreak}

\def\appendix#1#2{\bigskip\noindent%
\sectionfont \bf Appendix #1.\ #2
\nobreak\medskip\rm\nobreak}


\newcount\nref
\global\nref = 1

\def\ref#1#2{\xdef #1{[\number\nref]}
\ifnum\nref = 1\global\xdef\therefs{\noindent[\number\nref] #2\ }
\else
\global\xdef\oldrefs{\therefs}
\global\xdef\therefs{\oldrefs\vskip.1in\noindent[\number\nref] #2\ }%
\fi%
\global\advance\nref by 1
}

\def\listrefs{\vfill\eject\section{References}\therefs}


\newcount\nfig
\global\nfig = 1

\def\fg#1\efig{\vskip .5in\noindent Fig.\ \number\nfig:\ #1%
\global\advance\nfig by 1}


\newcount\cflag
\newcount\nequation
\global\nequation = 1
\def\eqlabel{(1)}

\def\nexteqno{\ifnum\cflag = 0
\global\advance\nequation by 1
\fi
\global\cflag = 0
\xdef\eqlabel{(\number\nequation)}}

\def\lasteqno{\global\advance\nequation by -1
\xdef\eqlabel{(\number\nequation)}}

\def\label#1{\xdef #1{(\number\nequation)}
\ifnum\dflag = 1
{\escapechar = -1
\xdef\draftname{\teenyfont\string#1}}
\fi}

\def\clabel#1#2{\xdef\eqlabel{(\number\nequation #2)}
\global\cflag = 1
\xdef #1{\eqlabel}
\ifnum\dflag = 1
{\escapechar = -1
\xdef\draftname{\string#1}}
\fi}

\def\cclabel#1#2{\xdef\eqlabel{#2)}
\global\cflag = 1
\xdef #1{\eqlabel}
\ifnum\dflag = 1
{\escapechar = -1
\xdef\draftname{\string#1}}
\fi}


\def\eeq{}

\def\eqnn #1\eeq{$$ #1 $$}

\def\eq #1\eeq{\xdef\draftname{\ }
$$ #1
\eqno{\eqlabel \rlap{\ \draftname}} $$
\nexteqno}







\def\eqa #1\eeq{\xdef\draftname{\ }
$$ \eqalignno{ #1 } $$
\global\cflag = 0}




\def\jref#1#2#3#4{{\it #1} {\bf #2}, #3 (#4)}

\def\NPB#1#2#3{\jref{Nucl.\ Phys.}{B#1}{#2}{#3}}

\def\PLB#1#2#3{\jref{Phys.\ Lett.}{#1B}{#2}{#3}}

\def\PRD#1#2#3{\jref{Phys.\ Rev.}{D#1}{#2}{#3}}

\def\PRL#1#2#3{\jref{Phys.\ Rev.\ Lett.}{#1}{#2}{#3}}


\def\goto{\mathop{\rightarrow}}


\def\frac#1#2{{{#1} \over {#2}}\,}  
\def\sfrac#1#2{{\textstyle\frac{#1}{#2}}}  

\def\grad{\vec\nabla}
\def\Dsl{\hbox{/\kern-.6000em\rm D}} 



\def\twi{\widetilde}

\def\scr#1{{\cal #1}}

\def\mybar#1{\kern 0.8pt\overline{\kern -0.8pt#1\kern -0.8pt}\kern 0.8pt}
\def\sla#1{\raise.15ex\hbox{$/$}\kern-.57em #1}
\def\Sla#1{\kern.15em\raise.15ex\hbox{$/$}\kern-.72em #1}

\def\roughly#1{\mathrel{\raise.3ex\hbox{$#1$\kern-.75em%
    \lower1ex\hbox{$\sim$}}}}


\def\tr{\mathop{\rm tr}}


\def\bra#1{\langle #1 |}
\def\ket#1{| #1 \rangle}



\def\al{\alpha}
\def\del{\delta}
\def\Del{\Delta}

\def\ep{\epsilon}
\def\lam{\lambda}
\def\Lam{\Lambda}
\def\om{\omega}

\def\sig{\sigma}

\def\CPT{\raise.45ex\hbox{$\chi$}PT}


\def\GeV{{\rm \ GeV}}



\def\qq{$Q\mybar Q$}


\id
LBL-33946, UCB-PTH-93/12
\endid

\title
\centerline{Quarkonium Decays and}
\centerline{Light Quark Masses}
\endtitle

\authors
\centerline{Markus A. Luty$^*$\ \ {\it and}\ \ Raman
Sundrum$^{*\dagger}$}
\vskip .1in
\centerline{\it $^*\,$Theoretical Physics Group}
\centerline{\it Lawrence Berkeley Laboratory}
\centerline{\it 1 Cyclotron Road}
\centerline{\it Berkeley, California 94720}\vskip .1in
\vskip .05in
\centerline{\it $^\dagger\,$Department of Physics}
\centerline{\it University of California}
\centerline{\it Berkeley, California 94720}
\footnote{}
{\hskip-.26in
This work was supported in part by the Director, Office of Energy
Research, Office of High Energy and Nuclear Physics, Division of High
Energy
Physics of the U.S. Department of Energy under contract
DE-AC03-76SF00098 and
in part by the National Science Foundation under grant PHY90-21139.}
\endauthors

\abstract
The $SU(3)$-violating decays $\Phi^{2S} \goto \Phi^{1S} X$, where
$X = \pi^0$ or $\eta$ and $\Phi = J/\psi$ or $\Upsilon$ have been
recently
proposed as a means of probing the light quark masses beyond leading
order
in chiral perturbation theory.
We argue that this analysis is incorrect, even in the heavy quark
limit.
We show that these decays are governed by an infinite number of
matrix
elements which are not suppressed by any small parameter, and which
cannot
be computed with our present understanding of QCD.
Furthermore, for sufficiently heavy quarks, we show that the decay
amplitudes
can be organized into a twist expansion, and that the contributions
considered in the above proposal are subleading in this expansion.
We also explain how these decays nonetheless give a constraint on the
light
quark masses valid at {\it leading order} in the chiral expansion.
The decays $\Phi^{1S} \goto \eta\gamma$ and $\Phi^{2S} \goto
\Phi^{1S} \pi\pi$
also have contributions from infinitely many operators, contrary to
claims in
the literature.
\endabstract


\ref\ismass{B. L. Ioffe and M. A. Shifman, \PLB{95}{99}{1980}.}

\ref\dw{J. F. Donoghue and D. Wyler, \PRD{45}{892}{1992};
J. F. Donoghue, B. R. Holstein, and D. Wyler, \PRL{69}{3444}{1992}.}

\ref\idea{M. Voloshin and V. Zakharov, \PRL{45}{688}{1980}.}

\ref\QCDmult{K. Gottfried, \PRL{40}{598}{1978}.}

\ref\peskin{M. E. Peskin, \NPB{156}{365}{1979};
M. E. Peskin and G. Bhanot, \NPB{156}{391}{1979}.}

\ref\rad{V. A. Novikov, M. A. Shifman, A. I. Vainshtein, and V. I.
Zakharov,
\NPB{165}{55}{1980}.}

\ref\muzero{The debate over whether $m_u = 0$ is allowed is ongoing.
See  D. B. Kaplan and A. V. Manohar \PRL{56}{2004}{1986};
H. Leutwyler, \NPB{337}{108}{1990};
K. Choi, \NPB{383}{58}{1992};
see also ref.\ \dw.}

\ref\choi{K. Choi, \PLB{292}{159}{1992}.}


\section{Introduction}

Recently, there has been a renewed interest in the flavor
$SU(3)$-violating
decays
\eq
\label\thedecay
\Phi^{2S} \goto \Phi^{1S} \pi^0
\quad{\rm and}\quad
\Phi^{2S} \goto \Phi^{1S} \eta,
\eeq
where $\Phi = J/\psi$ or $\Upsilon$, as a probe for the light quark
masses beyond leading order in chiral perturbation theory \ismass\dw.
This work is based on ref.\ \idea, where it is argued that the QCD
multipole expansion \QCDmult\peskin\ can be used to show that
\eq
\label\pform
\scr A(\Phi^{2S} \goto \Phi^{1S} X) \propto
\bra X \tr G^{\mu\nu} \twi G_{\mu\nu} \ket 0,
\eeq
where $X = \pi^0$ or $\eta$.
The basic idea in ref.\ \idea\ is that the $\Phi$ is spatially small
compared to the wavelength of gluons which are important in the decay
process, and that the decay can be computed in terms of the matrix
elements of local gluonic operators.
This result is supposed to be valid for sufficiently heavy quark masses.
The hadronic matrix element in eq.\ \pform\ can then be computed
using chiral perturbation theory \rad\dw.

In this paper, we show that eq.\ \pform\ is not the leading
contribution to the decays in eq.\ \thedecay:
there are an {\it infinite} number of operators which contribute whose
matrix elements are not suppressed by any small parameter.
This result is a simple consequence of the non-locality of the transition
amplitude {\it in time}.
Furthermore, in the heavy quark limit, the decay amplitude can be
organized into a twist expansion, and the contribution leading to
eq.\ \pform\ can be shown to be subleading in this expansion.
This invalidates the claim of ref.\ \dw\ that the decays in
eq.\ \thedecay\ can be used to constrain the light quark masses at
next-to-leading order in the chiral expansion.
Despite this negative result, we show that the decays \thedecay\ can be
used to derive a constraint on the light quark masses valid at
leading order in the chiral expansion, but not at higher orders.
This distinction is important, since the corrections to lowest-order
results are substantial, and may even allow $m_u = 0$, giving an
economical solution of the strong $CP$ problem \muzero.

A similar analysis can be used to show that the decays
\eq
\Phi^{2S} \goto \Phi^{1S} \pi\pi
\qquad{\rm and}\qquad
\Phi^{1S} \goto \eta\gamma
\eeq
also get contributions from infinitely many operators, contrary to
the claims of refs.\ \idea\ and \rad, respectively.

\section{Quantum Mechanics Analysis}

In this section, we adopt the assumptions and methodology used in
ref.\ \idea\ to describe the decay in eq.\ \thedecay.
We show that these assumptions lead to the conclusion that these
decays
are governed by an infinite number of matrix elements which are
unsuppressed by any small parameter.
In the next section, we will argue that this conclusion is much more
general than the argument of this section, and that the contributions
considered in this section are in fact not the dominant ones.

We will treat the $\Phi$ as a non-relativistic quantum-mechanical
bound state.
We begin by making the standard assumption that the decay in
eq.\ \thedecay\ can be viewed as a two-step process:
First the $\Phi^{2S}$ decays via the emission of two gluons, and then
the gluons ``hadronize'' to form the light pseudoscalar in the final
state.
The wavelength of the gluons involved is large compared to the spatial
size of the $\Phi$, and so gluon emission can be described in terms
of a local perturbation hamiltonian involving the gluon fields.

In ref.\ \idea, it is assumed that the leading
contribution to the decays in eq.\ \thedecay\ comes from interference
between the electric dipole interaction
\eq
\label\ed
\del H_1 = -g \vec r \cdot \vec E,
\eeq
and the the spin-dependent interaction
\eq
\label\spin
\del H_2 = -\frac g{2m_Q} (\vec r \cdot \vec D) (\vec S \cdot \vec
B),
\eeq
which arises at higher order in the nonrelativistic expansion.
Here, $\vec r$ is the separation between the quark and the antiquark,
$\vec S$ is the total spin operator, and
\eq
\label\ebform
\vec E \equiv -\grad A^0_a T_a, \qquad
\vec B \equiv \grad \times \vec A_a T_a,
\eeq
where
\eq
T_a \equiv \sfrac 12 (t_a \otimes 1 + 1 \otimes t_a^*).
\eeq

The matrix element is then given in second-order perturbation theory
by
\eq
\scr A \propto \bra f \del H_1 \,
\frac 1{\ep_i - (H_{Q\mybar Q} + H_g) + i0+} \, \del H_2 \ket i
+ (1 \leftrightarrow 2),
\eeq
where $\ep_i$ is the energy of the state $\ket i$, $H_{Q\mybar Q}$ is
the
hamiltonian of the \qq\ pair in the
QCD-generated potential and $H_g$ is the hamiltonian for free gluons.
Inserting a complete set of energy eigenstates, we obtain
\eq
\scr A \propto \sum_\al \bra f \del H_1 \,
\frac 1{\ep_i - H_{Q\mybar Q} - \om_\al + i0+}
\ket\al\bra\al \del H_2 \ket i + (1 \leftrightarrow 2),
\eeq
where $\om_\al$ is the energy carried away by the emission of the
first gluon.
We can expand this to obtain
\eq
\eqalign{
\scr A &\propto \sum_n \sum_\al \bra f \del H_1 \,
\frac 1{(\ep_i - H_{Q\mybar Q} + i0+)^{n + 1}}
\ket\al\bra\al \om_\al^n \del H_2 \ket i + (1 \leftrightarrow 2) \cr
&= \sum_n \bra f \del H_1 \,
\frac 1{(\ep_i - H_{Q\mybar Q} + i0+)^{n + 1}}
(i\partial_0)^n \del H_2 \ket i + (1 \leftrightarrow 2), \cr}
\eeq
where the time derivative $\partial_0$ is understood to act on the
gluon fields in $\del H_2$.
We therefore obtain
\eq
\label\pope
\eqalign{
\scr A(\Phi^{2S} \goto \Phi^{1S} X) &\propto
\frac {g^2}{2m_Q}\sum_n \bra{\Phi^{1S}} r_j \,
\frac 1{(\ep_i - H_{Q\mybar Q} + i0+)^{n + 1}} r_k S_\ell
\ket{\Phi^{2S}} \cr
&\qquad\qquad\qquad
\times \bra X E_j (iD_0)^n D_k B_\ell
+ D_j B_\ell (iD_0)^n E_k \ket 0, \cr}
\eeq
where we have included additional terms mandated by gauge invariance
by substituting the gauge-covariant derivative $D_0$ for the time
derivative
$\partial_0$.
Similar results are derived using field-theoretic arguments in
ref.\ \peskin.

If we expand this result in powers of $H_{Q\mybar Q}$, eq.\ \pope\ has the
form of an operator product expansion, with the suppression scale of the
higher-dimension operators given by the spacing between excited energy
levels $\Del$.
If we keep only the first term in this expansion, we recover the result
eq.\ \pform\ of ref.\ \idea.
However, the kinematic energy scale for these decays is $M_{2S} - M_{1S}$,
which is of order $\Del$, so there is no small parameter suppressing the
higher order terms in the series in eq.\ \pope, even in the heavy-quark
limit.
This is the central point of this paper.

To show explicitly that there is no suppression of higher-order terms
in eq.\ \pope, we follow ref.\ \idea\ and assume that the spatial and spin
parts of the $\Phi$ wavefunction factorize.
We can then write eq.\ \pope\ as
\eq
\label\main
\eqalign{
\scr A(\Phi^{2S} \goto \Phi^{1S} X) &\propto
\frac{g^2}{2m_Q} \sum_n \frac{c_n}{\Del^{n + 3}}\,
(\vec\Phi^{1S} \times \vec\Phi^{2S})_j \cr
&\qquad\qquad\qquad
\times \bra X E_k (iD_0)^n D_k B_j + D_k B_j (iD_0)^n E_k \ket 0,
\cr}
\eeq
where $c_n$ is a quantum-mechanical matrix element, and $\vec\Phi^{1S}$
and $\vec\Phi^{2S}$ are polarization vectors for the initial and final
quarkonium states.
In order to estimate the size of the hadronic matrix element, we write the
operator in relativistic notation using the four-velocity $v$ of the decaying
particle:
\eq
\label\amp
\scr A(\Phi^{2S} \goto \Phi^{1S} X) \propto
\frac{g^2}{2m_Q} \sum_n \frac{c_n}{\Del^{n + 3}}\,
v_\mu \ep^{\mu\nu\lam\rho} \Phi^{1S}_\nu \Phi^{2S}_\lam
v^\mu v^\nu v^{\lam_1} \cdots v^{\lam_n}
\bra X \scr O^{(n)}_{\rho\mu\nu\lam_1 \cdots \lam_n} \ket 0,
\eeq
where
\eq
\scr O^{(n)}_{\rho\mu\nu\lam_1 \cdots \lam_n} \equiv
\tr G_{\sig(\mu} D_{\lam_1} \cdots D_{\lam_n} D^\sig \twi
G_{\nu)\rho}
+ \tr D^\sig \twi G_{\rho(\mu} D_{\lam_1} \cdots D_{\lam_n}
G_{\nu)\sig}.
\eeq
Here, we use the notation $(\cdots)$ to denote the symmetrization of indices.
The hadronic matrix elements can be written
\eq
\label\vzform
\eqalign{
\bra X \scr O^{(n)}_{\rho\mu\nu\lam_1 \cdots \lam_n} \ket 0 &=
\Lam^2 \left[ g_{(\mu\nu} p_{\lam_1} \cdots p_{\lam_n)} p_\rho
- (\nu \leftrightarrow \rho) \right] F_1^{(n)} \cr
&\qquad
+ \Lam^4 \left[ g_{(\mu\nu} g_{\lam_1 \lam_2} p_{\lam_3}
\cdots p_{\lam_n)} p_\rho
- (\nu \leftrightarrow \rho) \right] F_2^{(n)} + \cdots, \cr}
\eeq
where $p$ is the momentum of $X$.
We have explicitly factored out powers of the hadronic scale $\Lam$ so that
all the form factors $F_j$ have the same dimension.
Since the form factors depend only on $p^2 = m_\pi^2$, they are constants
determined by low-energy QCD dynamics, and are all expected to be of the same
order of magnitude.
(Note that since the decay violates $SU(3)$, the form factors $F_j$ must be
proportional to quark masses.)

Substituting this into eq.\ \amp, we obtain
\eq
\label\relamp
\scr A(\Phi^{2S} \goto \Phi^{1S} X) \propto
\frac{g^2}{2m_Q}\, \Lam^2 \sum_n \frac{c_n}{\Del^{n + 3}}\,
v_\mu \ep^{\mu\nu\lam\rho} \Phi^{1S}_\nu \Phi^{2S}_\lam p_\rho\,\,
(p\cdot v)^n\,F_1^{(n)} + \cdots.
\eeq
Since $p\cdot v \sim \Del$, we see that the contributions from the
higher order terms in this expansion are unsuppressed, even in the
heavy quark limit.

Similar arguments which led to eq.\ \pform\ were used in ref.\ \idea\ to
argue that
\eq
\scr A(\Phi^{2S} \goto \Phi^{1S} \pi\pi) \propto
\bra{\pi\pi} \tr G^{\mu\nu} G_{\mu\nu} \ket 0,
\eeq
and in ref.\ \rad\ to argue that
\eq
\scr A(\Phi^{1S} \goto \eta\gamma) \propto
\bra{\eta} \tr G^{\mu\nu} \twi G_{\mu\nu} \ket 0.
\eeq
The considerations above can be used to show that these decays also have
unsuppressed contributions from an infinite number of operators.

\section{Effective Field Theory Analysis}

The conclusions of the previous section are not at all unexpected if we
consider these decays in an effective field theory language.
If we consider interactions of the $\Phi^{1S}$ state with gluons with
energies $E \ll \Del$, then we can write an effective field theory in
which the $\Phi^{1S}$ is treated as a heavy particle.
Higher-dimension terms in this effective lagrangian are suppressed by
powers of $\Del$, the energy required to excite the next excited state.
In order to consider decays such as those in eq.\ \thedecay, we must
write an effective theory containing fields for both $\Phi^{1S}$
and $\Phi^{2S}$ states interacting with gluons.
This is not a very useful effective theory, since the higher-dimension terms
in this lagrangian are again suppressed by powers of $\Del$, which is
also the kinematic scale for the decays we are interested in.
Therefore, we expect that an infinite number of gluonic operators will
contribute at the same level to the decay.
This situation arises because we are trying to describe a process which
is nonlocal (in time) on a scale $\sim 1 / \Del$ by local operators.

In an effective field theory for $\Phi$ radial transitions, we could
immediately write down contributions to the decay such as
eq.\ \amp\ and reach the conclusions of the previous section.
But we can also write terms such as
\eq
\label\theterm
\scr L_{\rm eff} = \frac{c}{\Del^4} v_\mu \ep^{\mu\nu\lam\rho}
\Phi^{1S}_\nu \Phi^{2S}_\lam
\tr(G^{\sig}_{(\al} D_{\lam_1} \cdots D_{\lam_n} D_\rho \twi G_{\beta)\sig})
v^\al v^\beta v^{\lam_1} \cdots v^{\lam_n} + \cdots.
\eeq
There is no symmetry which forbids these terms, so we expect them to be
generated at some level.
In fact, at higher orders in the non-relativistic expansion, factorization
of the spin and spatial parts of the $\Phi$ wavefunction assumed in
eq.\ \main\ breaks down, generating the contributions in eq.\ \theterm.
These terms give rise to contributions to the decay amplitude proportional to
\eq
\label\efform
\bra X \tr(G^{\sig}_{(\al} D_{\lam_1} \cdots D_{\lam_n} D_\rho \twi
 G_{\beta)\sig}) \ket 0
= p_\al p_\beta p_{\lam_1} \cdots p_{\lam_n} p_\rho G_1 + \cdots,
\eeq
where the form factor has been normalized to have the same dimension as
those in eq.\ \vzform.
In the heavy-quark limit, the contribution of this term to the decay
amplitude is enhanced relative to eq.\ \relamp\ by a factor of
\eq
g^m \left(\frac{p\cdot v}\Lam\right)^2 \sim
\left(\frac{m_Q}{\Lam}\right)^2,
\eeq
up to logarithms of $m_Q$, where $\Lam \sim 1 \GeV$ is a hadronic scale.

In general, for terms in the effective theory of the form
\eq
\scr L_{\rm eff} = \sum_n \frac{c_n}{\Del^n}\,
v_\mu \ep^{\mu\nu\lam\rho} \Phi^{1S}_\nu \Phi^{2S}_\lam
\scr O^{(n)}_\rho,
\eeq
where $\scr O^{(n)}$ is a QCD operator, it is easy to see that the leading
contributions will come from operators with lowest twist
($= {\rm dimension} - {\rm spin}$).
The operator in eq.\ \vzform\ has twist 4, while the operator in
eq.\ \efform\ has twist 2.
Thus, the contributions to the decay amplitude in  eq.\ \main\  are not
among the leading ones in the heavy-quark limit.

The relevance of the twist expansion to the real world is unclear, since
the $c$ and $b$ quarks are not sufficiently heavy to trust heavy-quark
results without reservation, and in any case the hadronic matrix elements
of the operators in question are unknown.

\section{Determination of the Light Quark Masses}

Our main motivation for this work is the fact that ref.\ \dw\ argued
that the decays in eq.\ \thedecay\ can be used to give a constraint
on the light quark masses beyond leading order in the chiral expansion.
We have argued that these decays are not dominated by the matrix elements
in eq.\ \pform, so it may seem mysterious that the constraint obtained
in ref.\ \dw\ is consistent with other determinations of the light quark
mass ratios.
However, it is easy to see that the decays \thedecay\ give a constraint
on the light quark masses valid at leading order in the chiral expansion.

The point is that the decay is forbidden in the limit of exact vector
$SU(3)$.
However, $SU(3)$ is broken explicitly by the quark masses, so the
leading contribution to the amplitude can be written
\eq
\scr A(\Phi^{2S} \goto \Phi^{1S} X) \propto \tr(M_q T_X),
\eeq
where $M_q$ is the light quark mass matrix and $T_X$ is the $SU(3)$
generator corresponding to the pseudoscalar in the final state.
(We can use {\it chiral} invariance to show that a similar term does
not exist due to electromagnetic breaking of $SU(3)$.)
This form of the amplitude immediately gives the leading order constraint
on the light quark masses used in ref.\ \dw.
Similar arguments to those given above can be found in ref.\ \ismass.

\section{Conclusions}

We have shown that the decays in eq.\ \thedecay\ cannot be used to
constrain the light quark masses beyond leading order.
These decays are governed by an infinite number of operators which are
unsuppressed by any small parameter even in the heavy quark limit.
Furthermore, for sufficiently heavy quark masses, the leading contributions
to the decay come from operators of twist 2, while the contribution
discussed in ref.\ \idea\ has twist 4.

The use of the decays in eq.\ \thedecay\ to determine the light quark
masses has also been criticized in ref.\ \choi.
However, this paper assumes that the leading contribution to the decay
comes from the first term of eq. \main, and criticizes the quantitative
accuracy of the heavy quark and $1 / N_{\rm color}$ expansion used in
ref. \idea.

\section{Acknowlegements}

We would like to thank M. E. Peskin and M. Suzuki for discussions.
This work was supported in part by the Director, Office of Energy Research,
Office of High Energy and Nuclear Physics, Division of High Energy Physics of
the U.S. Department of Energy under contract DE-AC03-76SF00098 and in part by
the National Science Foundation under grant PHY90-21139.

\listrefs
\bye